\documentclass[a4paper, 10 pt, conference]{cras} 
\usepackage[utf8]{inputenc}
\IEEEoverridecommandlockouts                       
\usepackage{mathtools}
\usepackage{geometry}
\usepackage{newtxmath,newtxtext}
\usepackage{authblk}
\usepackage{pgfplots}
\usepackage{todonotes}
\usepackage{subfig}
\usepackage{graphicx}
\usepackage{svg}
\usepackage[colorlinks,urlcolor=blue]{hyperref} 
\pgfplotsset{compat=newest} 
\pgfplotsset{plot coordinates/math parser=false}

\title{\Large \bf
VideoSum: A Python Library for Surgical Video Summarization
}

\author{\large Luis C. Garcia-Peraza-Herrera}
\author{\large S\'ebastien Ourselin} 
\author{\large Tom Vercauteren} 
\vspace{10pt}
\affil{\small\textit{King's College London}}

\begin{document}
\bstctlcite{IEEEexample:BSTcontrol}

\maketitle
\thispagestyle{empty}
\pagestyle{empty}

\section*{INTRODUCTION}
The performance of deep learning (DL) algorithms is heavily influenced by the quantity and quality of the annotated data. However, in Surgical Data Science (SDS)~\cite{Maier-Hein2022}, access to it is limited.
It is thus unsurprising that substantial
research efforts are made to develop methods aiming at mitigating the scarcity of annotated SDS data.
In parallel, an increasing number of Computer Assisted Interventions (CAI) datasets are being 
released\footnote{See \url{https://github.com/luiscarlosgph/list-of-surgical-tool-datasets}},
although the scale of these remains limited.
On these premises, data curation is becoming a key element of many SDS research endeavors.

Surgical video datasets are demanding to curate and would benefit from dedicated support tools.
With an average duration of $130.45$ minutes \cite{Costa2017}, surgical interventions pose challenges in terms of visualization, annotation and processing.

In this work, we propose to summarize surgical videos into \textit{storyboards} or \textit{collages} of representative frames to ease visualization, annotation, and processing.
Video summarization is well-established for natural images \cite{Apostolidis2021}.
However, state-of-the-art methods typically rely on models trained on human-made annotations \cite{Apostolidis2021a},
few methods have been evaluated on surgical videos and the availability of software packages for the task is limited. 
We present \texttt{videosum}\footnote{See
\url{https://github.com/luiscarlosgph/videosum}},
an easy-to-use and open-source Python library to generate \textit{storyboards} from surgical videos that contains a variety of unsupervised methods.
\begin{figure}[!t]
    \centering
	\includegraphics[width=.9\columnwidth]{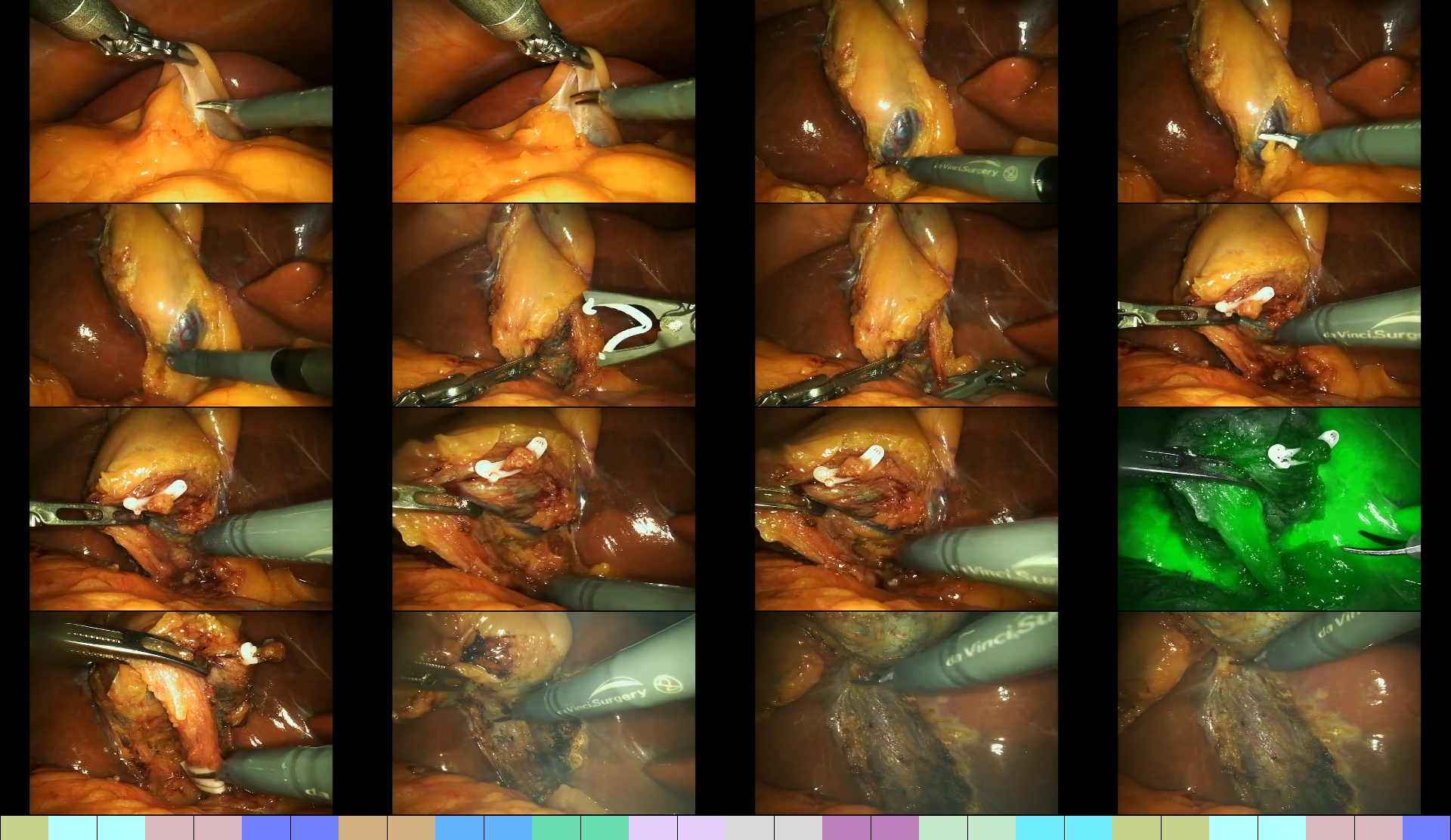}
	\caption{Baseline storyboard of a surgical video produced by the \textit{time} method of \texttt{videosum}.
    The bar under the collage represents the length of the video. 
    The colors represent the cluster label of the video frames, and the black vertical bars are the key frames.
    The video is split into uniform temporal segments by the \textit{time} method.
	}
	\label{fig:time}
\end{figure} 
\section*{MATERIALS AND METHODS}
Although there are many ways one could opt for to summarize a surgical video, we
convert each video into a \textit{storyboard}, i.e. a set of representative video frames,
as the one shown in Fig.~\ref{fig:time}. 
Formally, we aim to solve for a mapping $f$ such that 
$f_{\boldsymbol{\theta}}(X) \approx \{ Y, Z\}$, 
where 
$f_\mathbf{\theta}$ is a parametric function,
$\boldsymbol{\theta}$ is a vector of parameters,
$X = \{ \boldsymbol{x}_i \}_{i = 1}^{N_f}$ is a set of ordered video frames $\boldsymbol{x}_i \in \mathbb{R}^{H\times W \times C}$,
$N_f \in \mathbb{N}$ is the number of frames in the video $X$,
$Y = \{ y_i \}_{i = 1}^{N_c}$ is a set containing the indices $\boldsymbol{y}_i \in \{1, ..., N_f \}$ of the key frames,
$N_c \in \mathbb{N}$ is the number of clusters to be included in the storyboard,
and
$Z = \{ z_i \}_{i = 1}^{N_f}$ is an ordered set of clustering labels 
$z_i \in \{1, ..., N_c\}$
indicating to which cluster an input frame belongs.

At present, four different methods are available in
\texttt{videosum}:
\textit{time}, \textit{inception}, \textit{uid}
\footnote{\textit{uid} stands for univariate inception distance.}, 
and \textit{scda} \cite{Wei2017a}. 
The \textit{time} method trivially partitions the video in equal temporal parts and selects the median frame index of each cluster as key frame.
In all the other methods, the frames of the video are clustered using $k$-medoids. 
The difference between \textit{inception}, \textit{uid}, and \textit{scda} lies on two aspects:
1) how the feature vector is computed for any given frame;
and 2) what distance metric is used for clustering.
In the \textit{inception} method, the
representation of a frame is 
provided by
the $2048$-element 
latent
vector
from
InceptionV3 \cite{Szegedy2016} trained on ImageNet. The distance metric for clustering
in this case is the $\ell^2$-norm.
In the \textit{uid} method, InceptionV3 latent vectors are also used, but the clustering metric
is the 2-Wasserstein distance computed between univariate Gaussian models fitted on the elements of the inception feature vector.
With $\mu_i$ the mean and $\sigma_i$ the standard deviation of the $2048$-element vector computed from frame $i$, the distance between frame $i$ and $j$ reads 
$\sqrt{(\mu_i - \mu_j)^2 + (\sigma_i - \sigma_j)^2}$.
In the \textit{scda} method, the feature vector of a frame is computed as proposed in~\cite{Wei2017a} (although we started from InceptionV3 instead of VGG-16), and the $\ell_2$-norm is used as a distance metric for clustering.
\iffalse
In \cite{Wei2017a}, the authors use the low-resolution latent tensor of any given model --VGG in their case-- as input.
The assumption is that if many channels fire at the same region, we could expect this region to be an object.
\fi
To generate the image descriptor, \cite{Wei2017a} use a low-resolution latent tensor in the network.
They sum it over channels and threshold the collapsed tensor to obtain a binary mask.
The largest connected component of this binary mask is used to select indices in the low-resolution tensor and create the representation vector for the input image.

One of the risks of selecting visually distinct frames is that storyboards could concentrate on a short part of the surgery. To mitigate it, \texttt{videosum} supports time smoothing, allowing the user to select how close the summary should be to the \textit{time} version.

\section*{RESULTS}
\begin{figure}[!t]
    \centering
	\includegraphics[width=.9\columnwidth]{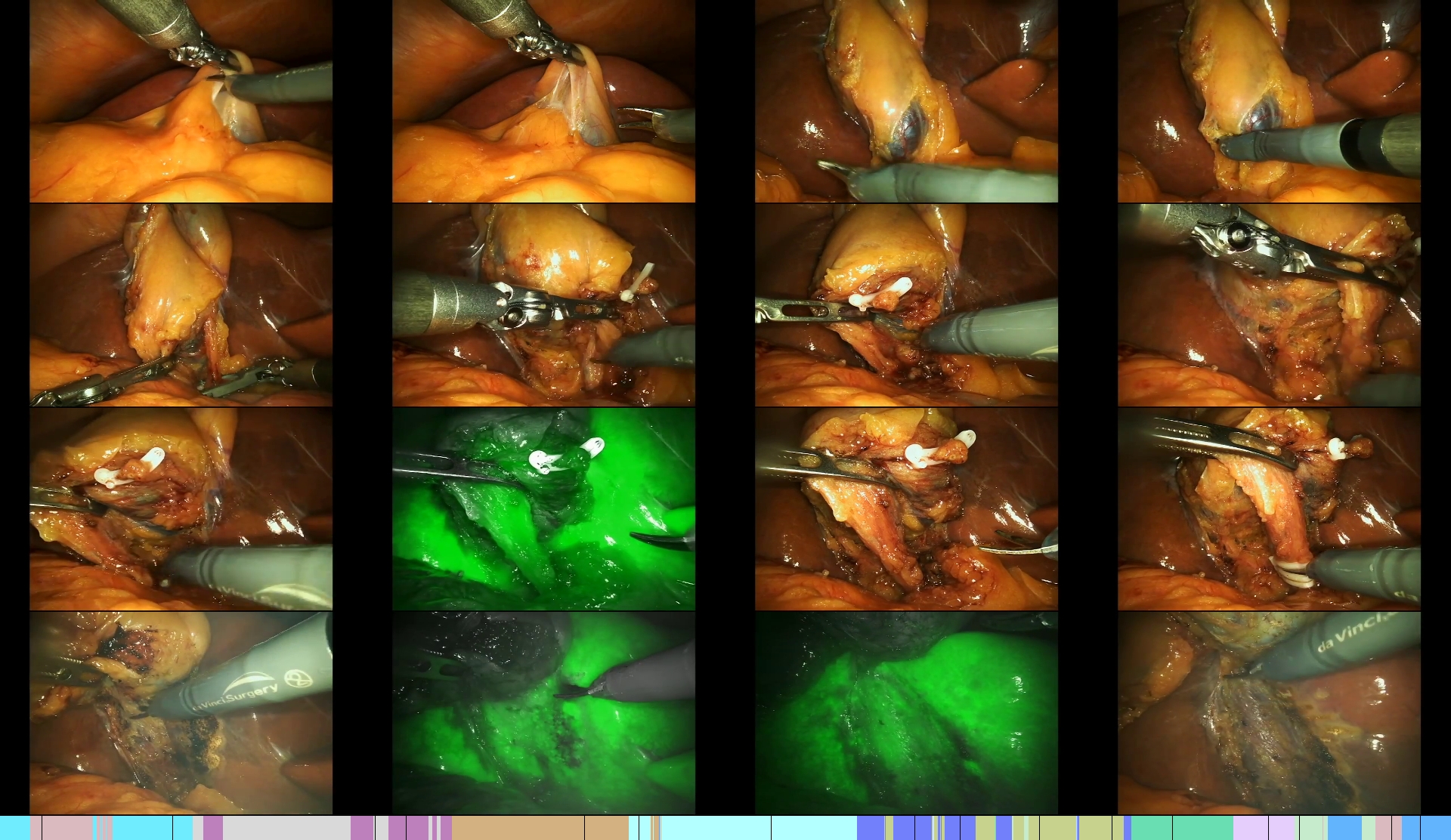}
	\caption{Storyboard produced by the \textit{inception} method.
	}
	\label{fig:inception}
\end{figure}
\begin{figure}[!t]
    \centering
	\includegraphics[width=.9\columnwidth]{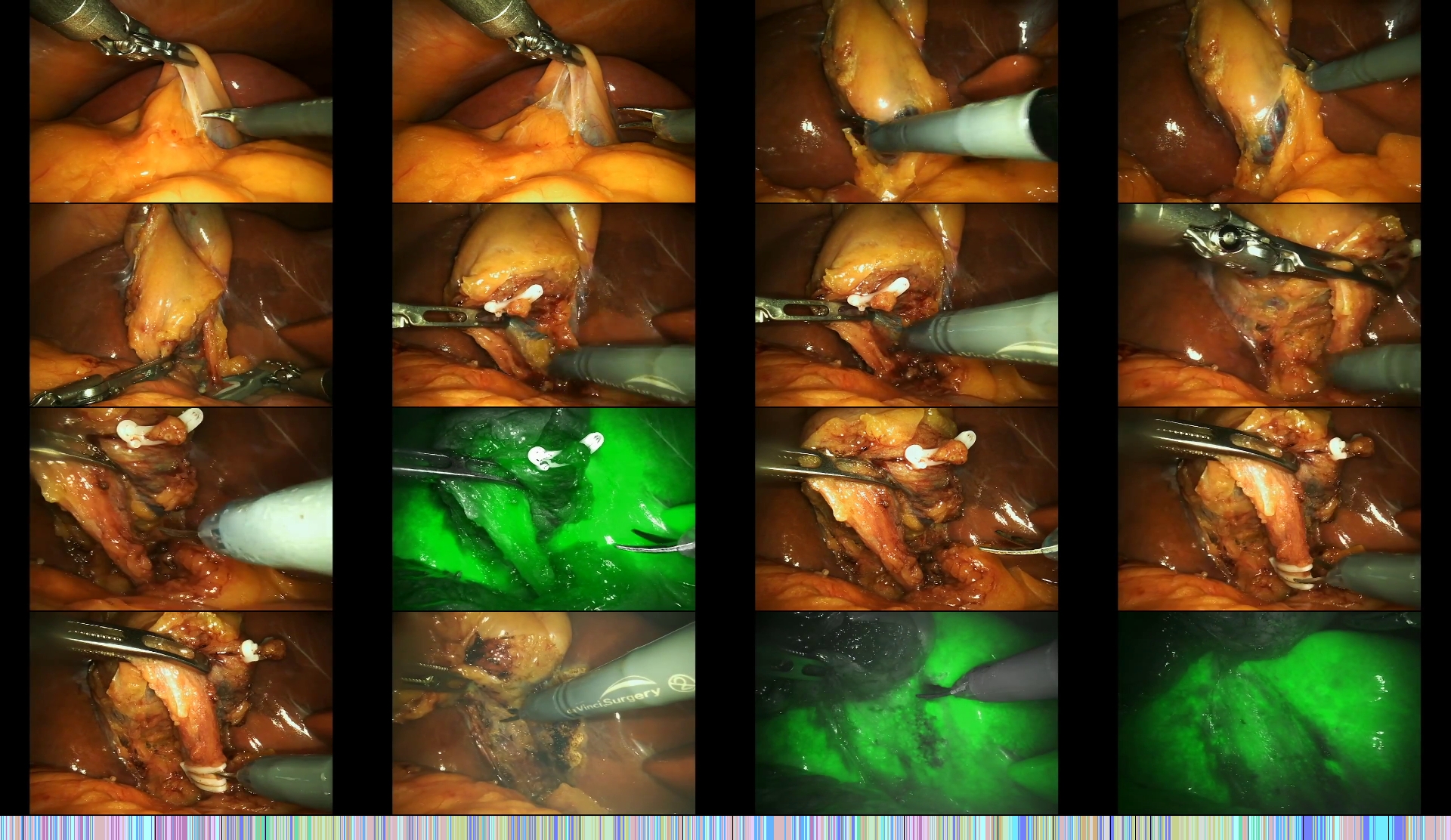}
	\caption{Storyboard produced by the \textit{uid} method.
	}
	\label{fig:uid}
\end{figure}
\begin{figure}[!t]
    \centering
	\includegraphics[width=.9\columnwidth]{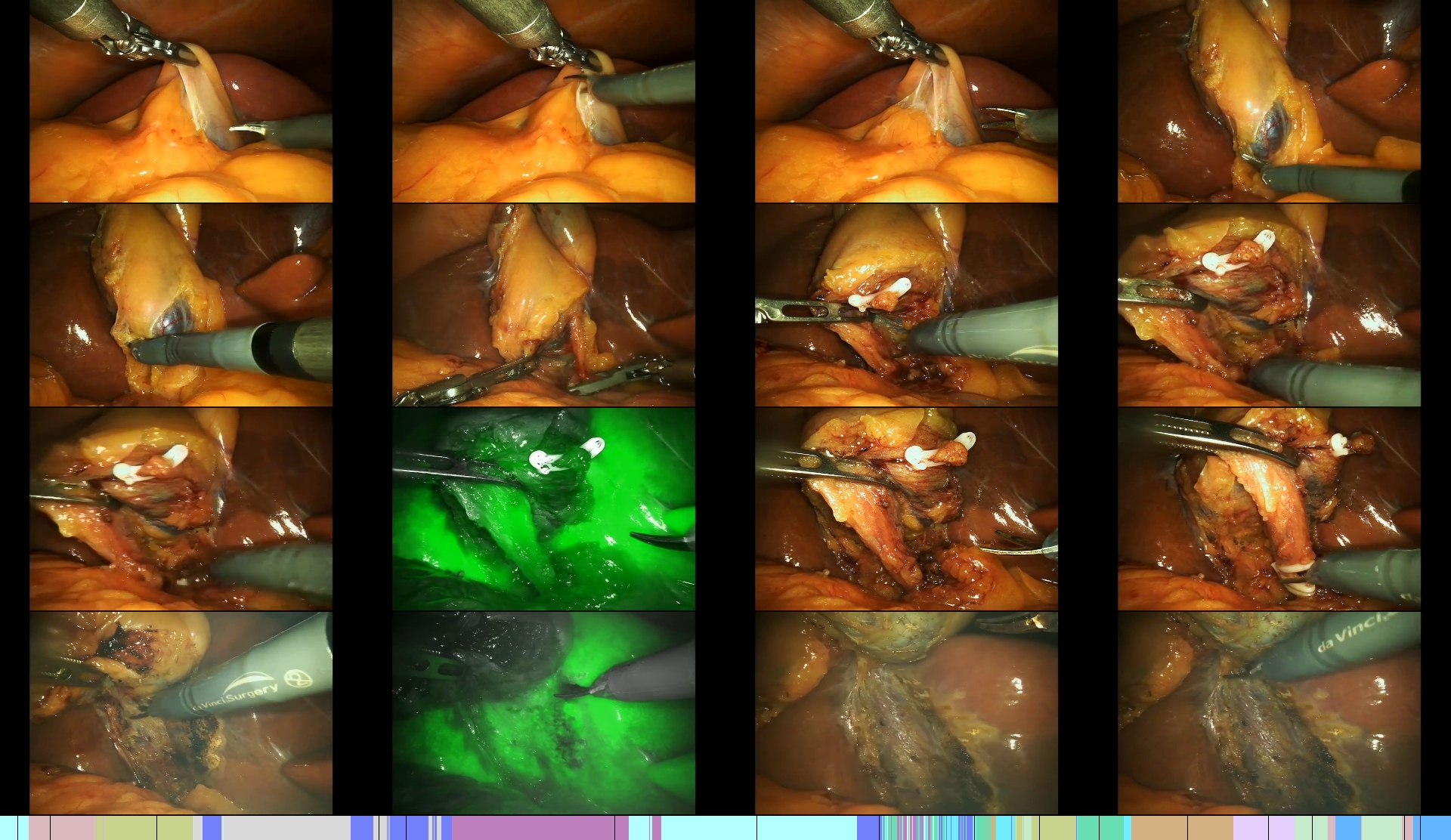}
	\caption{Storyboard produced by the \textit{scda} method.
	}
	\label{fig:scda}
\end{figure}

\begin{figure}[t]
\centering
\includegraphics[width=.9\columnwidth]{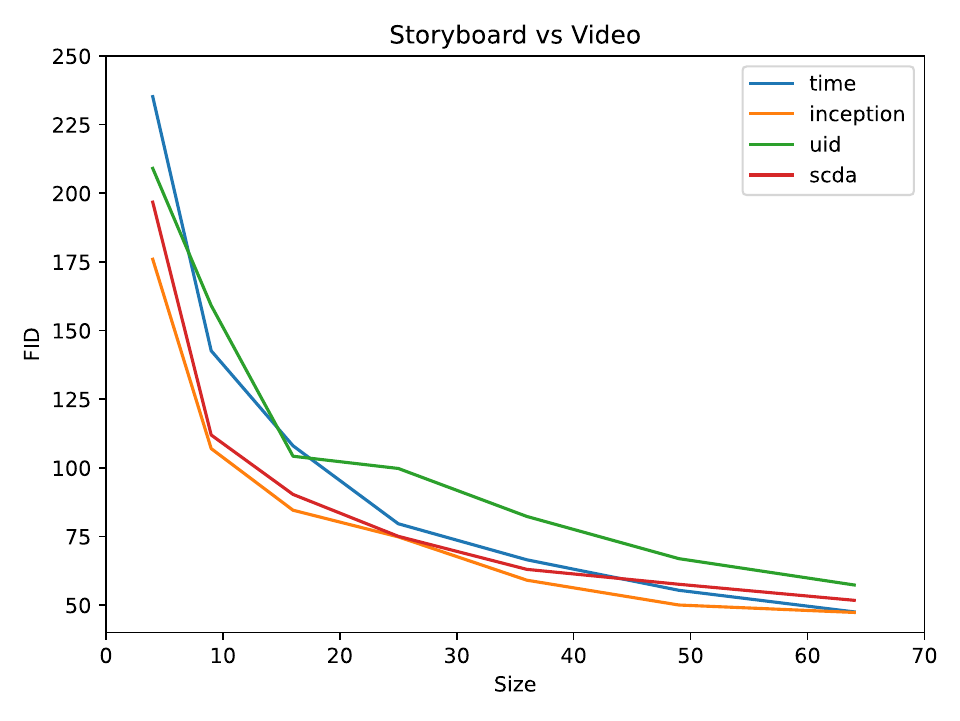}
\caption{Fr\'echet Inception Distance (FID) between the Gaussian distribution estimated from the frames in the storyboard and that obtained using all the frames in the video (lower is better, as it indicates that the storyboard is closer to the original video). The size indicates the number of key frames collected in the storyboard.
}
\label{fig:plot}
\end{figure}

An exemplary storyboard for the methods \textit{inception}, \textit{uid}, and \textit{scda} is shown in Fig. \ref{fig:inception}, \ref{fig:uid}, and \ref{fig:scda}, respectively. These storyboards correspond to the same video\footnote{See \url{https://github.com/luiscarlosgph/videosum/blob/main/test/data/video.mp4}} whose \textit{time} summary is presented in Fig. \ref{fig:time}.
In order to evaluate the different methods, we use the \textit{Fr\'echet Inception Distance} (FID) \cite{Heusel2017}. 
The FID compares two Gaussian distributions. 
To compare how close the storyboard is to the video, we estimate one Gaussian from the feature vectors of \textit{all} the frames of a video, and another one from the feature vectors of \textit{only} those frames included in the storyboard.
In Fig.~\ref{fig:plot}, we show how the FID changes (for the video shown in Fig. \ref{fig:time}) as we increase the number of images in the storyboard for each one of the methods.
As shown in Fig. \ref{fig:plot}, the \textit{inception} method outperforms all the others for collages of different sizes. 
Without sophisticated optimizations, the current run times for an hour of video sampled at $1\,$fps are $13\,$s, $86\,$s, $216\,s$, and $74\,$s for \textit{time}, \textit{inception}, \textit{uid}, and \textit{scda}, respectively.

\section*{DISCUSSION AND CONCLUSION}
As shown in Fig. \ref{fig:uid}, treating the latent vector of a frame as set of samples of a univariate Gaussian distribution (\textit{uid} method) leads to noisy clustering. In terms of running time, \textit{inception} and \textit{scda} are comparable, but the improved performance of \textit{inception} across a wide range of storyboard sizes makes it a strong candidate to serve as an unsupervised baseline for surgical video summarization. 
\section*{ACKNOWLEDGMENTS}
This project has received funding from the European Union’s Horizon 2020 research
and innovation programme under grant agreement No 101016985 (FAROS project). 
\bibliographystyle{IEEEtran}
\bibliography{references/library,references/ieeebib_settings}

\end{document}